# Beamlines and degraders for proton therapy

*A. Gerbershagen*
John Adams Institute, University of Oxford, Oxford, United Kingdom

**Abstract**
The accelerator in a proton therapy facility produces a beam, but the clinical treatment is defined by the beam delivery system. This note summarizes the beam transport and delivery systems between the medical accelerator and the patient. It briefly covers the physical motivation for proton therapy, focusing on the Bragg peak and the need to control range, transverse dose distribution, and irradiation time. The note then introduces magnetic rigidity, beam transport, degraders, energy selection systems, chromaticity, and dispersion suppression, with particular attention to fixed-energy cyclotron facilities in which treatment energies are obtained by controlled degradation. Gantry concepts are discussed from the viewpoint of beam optics, mechanical constraints, achromaticity, scanning geometry, and phase-space matching. The second part compares passive scattering with pencil-beam scanning and discusses upstream and downstream scanning, multiple Coulomb scattering, range shifters, treatment nozzles, organ-motion mitigation, volumetric rescanning, and large-momentum-acceptance transport.

## 1 Longitudinal dose and the Bragg peak

The clinical motivation for proton therapy is the finite range of charged particles in tissue. A proton beam deposits a moderate entrance dose, loses energy through electromagnetic interactions with atomic electrons, and reaches a maximum stopping power close to the end of its range. This maximum is the Bragg peak. The proton kinetic energy therefore determines the penetration depth, while the beam current and irradiation time determine the delivered dose [1,2].

A useful approximation for the proton CSDA range in liquid water is

$$R_{\text{water}} \approx 0.0022\ \text{cm}\ (T\,/\,\text{MeV})^{1.77}, \tag{1}$$

where T is the proton kinetic energy. For a fixed material composition, the physical range scales approximately as 1/ρ. In treatment planning, path lengths through different tissues are converted to water-equivalent thickness using relative stopping power rather than density alone.

As shown in Fig. 1, a monoenergetic Bragg peak is too narrow to cover most tumours longitudinally. Clinical dose delivery therefore superimposes many energies to form a spread-out Bragg peak (SOBP). The beam delivery system must generate these energies accurately, transport the particles to the patient, and distribute them transversely over the target volume.

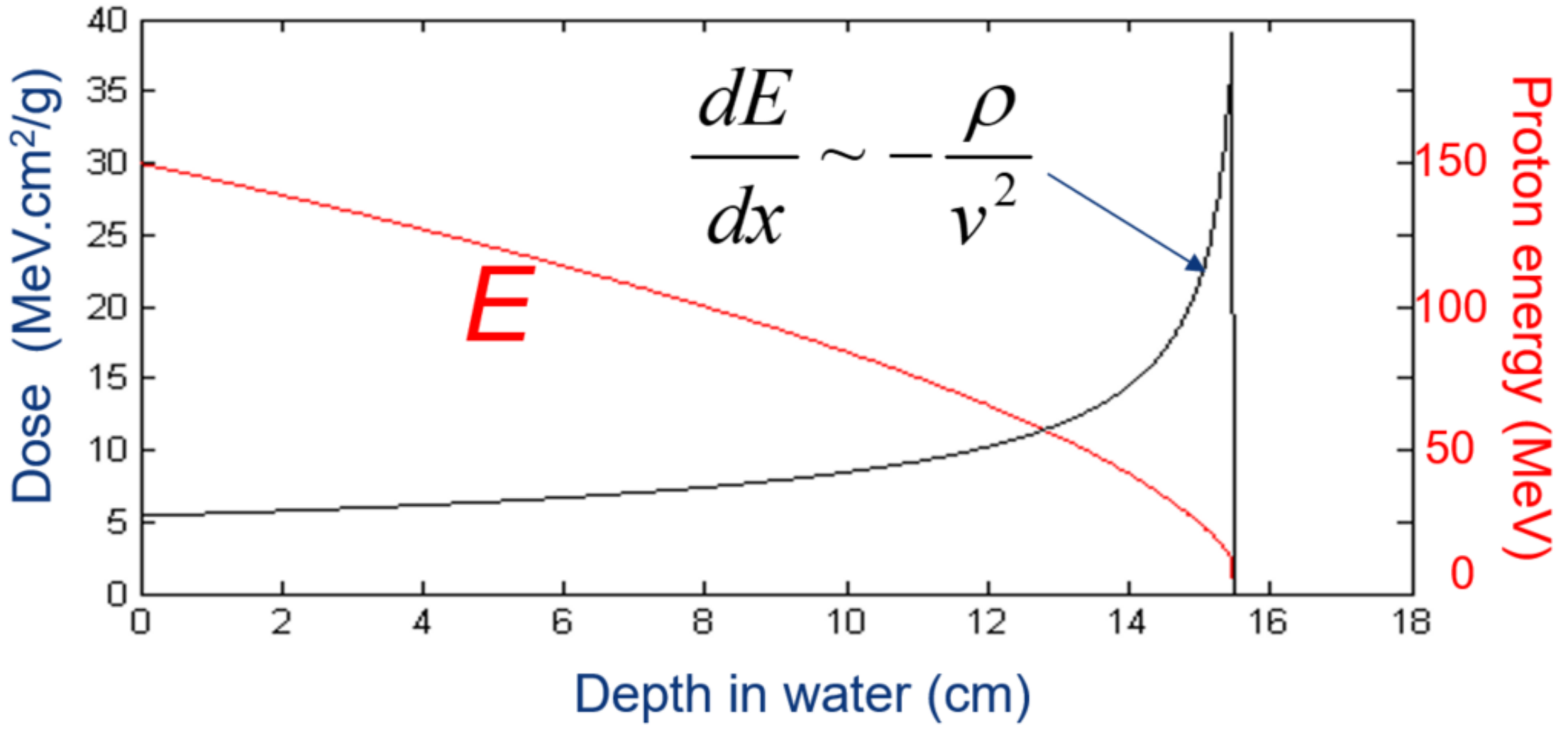

**Fig. 1:** Longitudinal dose deposition by protons. As the particles slow down, the stopping power rises and produces the Bragg peak. The beam energy determines the range in water.

## 2 Beam transport basics

After extraction from the accelerator, the beam is transported through dipole and quadrupole magnets. The dipoles define the central trajectory. The quadrupoles control the beam envelope and match the beam into downstream systems such as degraders, energy selection sections, scanning magnets, and treatment nozzles.

### 2.1 Magnetic rigidity, dipoles and quadrupoles

The magnetic force is perpendicular to the velocity. It changes the beam direction but does not change the kinetic energy. Equating the Lorentz force with the centripetal force gives the magnetic rigidity

$$B\rho = p\ /\ |q|. \tag{2}$$

In practical units,

$$B\rho\ [\mathrm{T\ m}] = 3.33564\ p\ [\mathrm{GeV/c}]\ /\ (|q|/e). \tag{3}$$

A 250 MeV proton has Bρ ≈ 2.43 T m. A fully stripped carbon ion at 450 MeV/u has Bρ ≈ 6.8 T m. This difference in rigidity explains much of the size difference between proton and ion beam transport systems. A quadrupole field increases linearly with displacement from the axis. It focuses in one transverse plane and defocuses in the other. Focusing in both planes therefore requires combinations of quadrupoles, for example doublets or triplets.

### 2.2 Phase space, emittance and envelopes

The transverse state of a particle can be described by position x and angle x′. The distribution of many particles forms an ellipse in phase space. In the usual Courant-Snyder convention its area is π ε, where ε is the geometric emittance. In linear, conservative optics the emittance is conserved. The beam size changes along the line because the ellipse rotates and is transformed by lenses and drifts. Material interactions and nonlinear fields can increase the rms emittance, while apertures truncate the transmitted phase-space distribution.

The envelope through a beamline is calculated from the initial Twiss parameters, the emittance, and the transfer matrices of all optical elements. At a beam waist, the correlation between position and angle vanishes. Upstream of the waist the beam is converging, implying a negative correlation between x and x′, and downstream it is diverging, giving a positive correlation. The three characteristic cases are shown in Fig. 2. These phase-space pictures explain why beam transport lines require alternating focusing and defocusing elements.

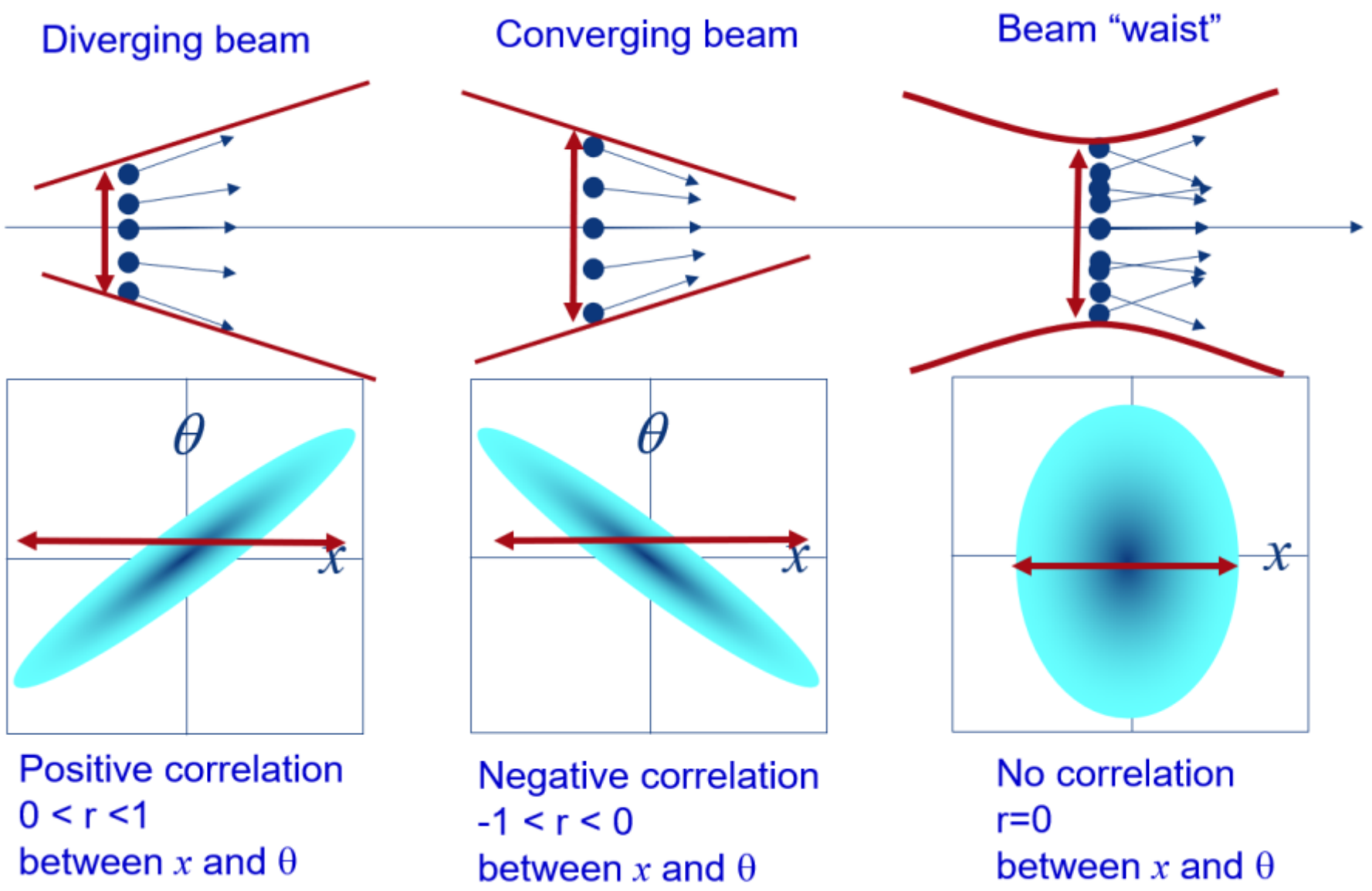

**Fig. 2:** Phase-space diagrams for a diverging beam (left), a converging beam (centre), and a beam waist (right).

### 2.3 Dispersion, achromaticity and imaging

Particles whose momentum differs from the nominal value follow different trajectories in bending magnets. This creates dispersion. In first-order notation, the horizontal position at a downstream point can be written as

$$x_2 = R_{11}\, x_1 + R_{12}\, x_1' + R_{16}\, \delta, \quad \delta = (p - p_0) / p_0. \tag{4}$$

The coefficient $R_{16}$ describes the horizontal position shift caused by a relative momentum error, while the corresponding output-angle dependence contains $R_{26}\delta$. In an uncoupled planar beamline, a first-order achromat therefore requires $R_{16} = R_{26} = 0$ at its end. If vertical dispersion or transverse coupling is present, $R_{36} = R_{46} = 0$ must also be satisfied. Achromaticity is required when momentum spread must not shift the beam spot at the patient or at a diagnostic point.

Imaging is another common optical condition. Point-to-point imaging from one location to another requires $R_{12} = 0$, so the final position is independent of the initial angle. Point-to-parallel imaging requires $R_{22} = 0$, so particles from one initial point emerge with the same angle. These conditions are useful for projecting apertures, defining beam sizes, and designing scanning systems.

## 3 Degraders and energy selection

The method used to set the beam energy depends on the accelerator type. Linacs can vary the final energy by changing the RF amplitude or by switching accelerating sections. Synchrotrons can extract at different energies, although changing energy often requires a new magnetic cycle. Most clinical cyclotrons extract at a fixed energy, usually around 230 to 250 MeV for proton therapy. The treatment energy must then be reduced downstream of the accelerator.

### 3.1 Degrader systems

A degrader places a controlled thickness of material in the beam path. The protons lose energy by ionization and emerge with a lower mean energy. In modern systems the degrader is often made from moving wedges, which permit rapid and reproducible changes in water-equivalent range. At PSI, the wedges can traverse the 230 to 70 MeV setting range in less than 50 ms; including beamline retuning, a typical 5 mm range step takes about 80 to 100 ms [3,7].

Energy degradation broadens the beam in both momentum and transverse phase space. Energy straggling increases the momentum spread, multiple Coulomb scattering increases the transverse emittance, and nuclear interactions produce secondary radiation. Collimators are therefore placed close to the degrader, where losses are expected and can be shielded. Transmission depends strongly on energy and beamline design. At low clinical energies it can be more than two orders of magnitude below that at 230 MeV [4,7].

### 3.2 Energy selection systems

The beam emerging from a degrader has a momentum spread too large for precise range control. An energy selection system reduces this momentum spread. A dispersive section correlates horizontal position with momentum. The optics are arranged so that the monochromatic transverse beam size is small at the selection slit while the dispersion is large. At this location, transverse position therefore serves as a measure of momentum.

A mechanical slit or collimator then transmits only the desired momentum interval. Typical accepted momentum intervals are of order ±1% or less, depending on the facility and treatment technique [4,5]. The selected beam is then transported through an achromatic section so that the residual momentum spread does not create an unwanted position offset downstream. The ESS is therefore both a beam-cleaning system and a range-definition system. The degrader and energy-selection sequence is illustrated in Fig. 3.

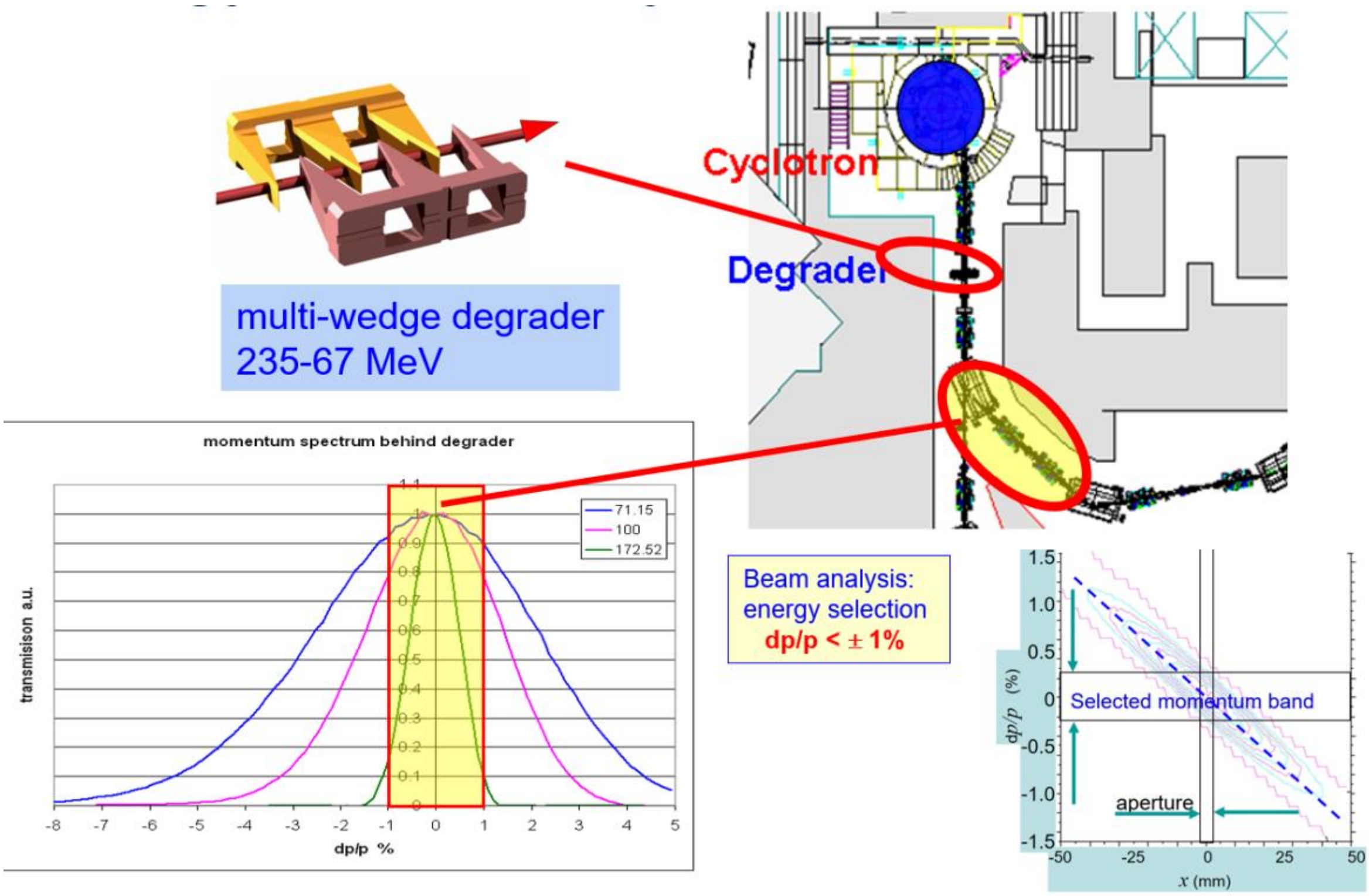

**Fig. 3:** Cyclotron-based proton therapy facility and energy-selection principle. The degrader (top left) broadens the momentum distribution (bottom left), and the energy selection system (bottom right) limits the accepted width before transport to the treatment rooms (top right).

# 4 Transverse dose delivery

Depth control alone does not define a treatment field, because the dose must also be distributed transversely over a target volume that is usually irregular. Two approaches are used:

- Passive scattering broadens the beam and shapes it with material devices.
- Pencil-beam scanning keeps the beam narrow and moves it magnetically from point to point.

## 4.1 Multiple Coulomb scattering

Beam broadening in matter is dominated by Coulomb scattering from nuclei, while the energy loss is dominated by interactions with atomic electrons. After many small-angle deflections, the central part of the projected angular distribution is well approximated by a Gaussian whose width is given by the Lynch-Dahl form of the Highland approximation [6].

$$\theta_0 \approx 13.6\ \text{MeV}\ z\ /\ (\beta\, p\, c)\ \sqrt{(x\ /\ X_0)}\ [1 + 0.038\ \ln(x\, z^2\ /\ (X_0\, \beta^2))]. \tag{5}$$

Here $\theta_0$ is the projected rms width of the Gaussian core, $x/X_0$ is the material thickness in radiation lengths, p is momentum, βc is velocity, and z is the projectile charge number. The important point for beamline design is that energy loss and scattering scale differently with material. A high-Z material produces much stronger angular scattering for similar energy loss. For 250 MeV protons, about 1 cm of water and 1 mm of tungsten both give an energy loss of roughly 4 MeV, but the characteristic scattering angles are about 5 mrad and 16 mrad, respectively.

## 4.2 Passive scattering

A passive scattering nozzle uses high-Z scatterers to broaden the beam. A single scatterer produces a broad but still non-uniform fluence. To improve the uniformity, double-scattering systems use shaped scatterers and compensating material to flatten the field while keeping the energy loss more uniform across the aperture. Patient-specific collimators or multi-leaf systems shape the lateral field.

Passive systems also contain range modulators, dose monitors, and range compensators. The compensator shapes the distal edge of the dose distribution to match the target depth. The aperture shapes the transverse field. The main disadvantage is the amount of material placed in the beam path. Scatterers, apertures, compensators, and air gaps increase the lateral penumbra. Interactions in collimators and compensators also generate secondary neutrons and activate nearby components [7].

## 4.3 Pencil-beam scanning

Pencil-beam scanning removes most passive material from the nozzle. Fast scanning magnets deflect a narrow beam across the treatment field. The vacuum chamber is extended as close as practical to the patient to reduce scattering in air. Range shifters are used mainly for shallow targets and are made from low-Z materials to minimize additional scattering.

Scanning paints the target in energy layers. The energy defines the depth layer, while the scanning magnets define the transverse position. The dose contribution of each spot can be optimized independently. This enables intensity-modulated proton therapy. Compared with passive scattering, pencil-beam scanning can improve proximal dose conformity, reduce the need for patient-specific hardware, and generally lower externally produced neutron dose [7,8]. Range shifters and collimators can reduce some of these advantages.

The most common clinical scanning mode is step-and-shoot spot scanning. The magnets move the beam to a spot position, the requested monitor units are delivered, the beam is switched off, and the magnets move to the next spot. A one-litre target sampled on roughly a 5 mm grid contains of order 10 000 spot positions, although the actual number depends on margins, layer spacing, and optimization.

The method is accurate, but treatment time accumulates through spot settling, beam switching, dose verification, and energy-layer switching. The corresponding layer-painting pattern is shown in Fig. 4.

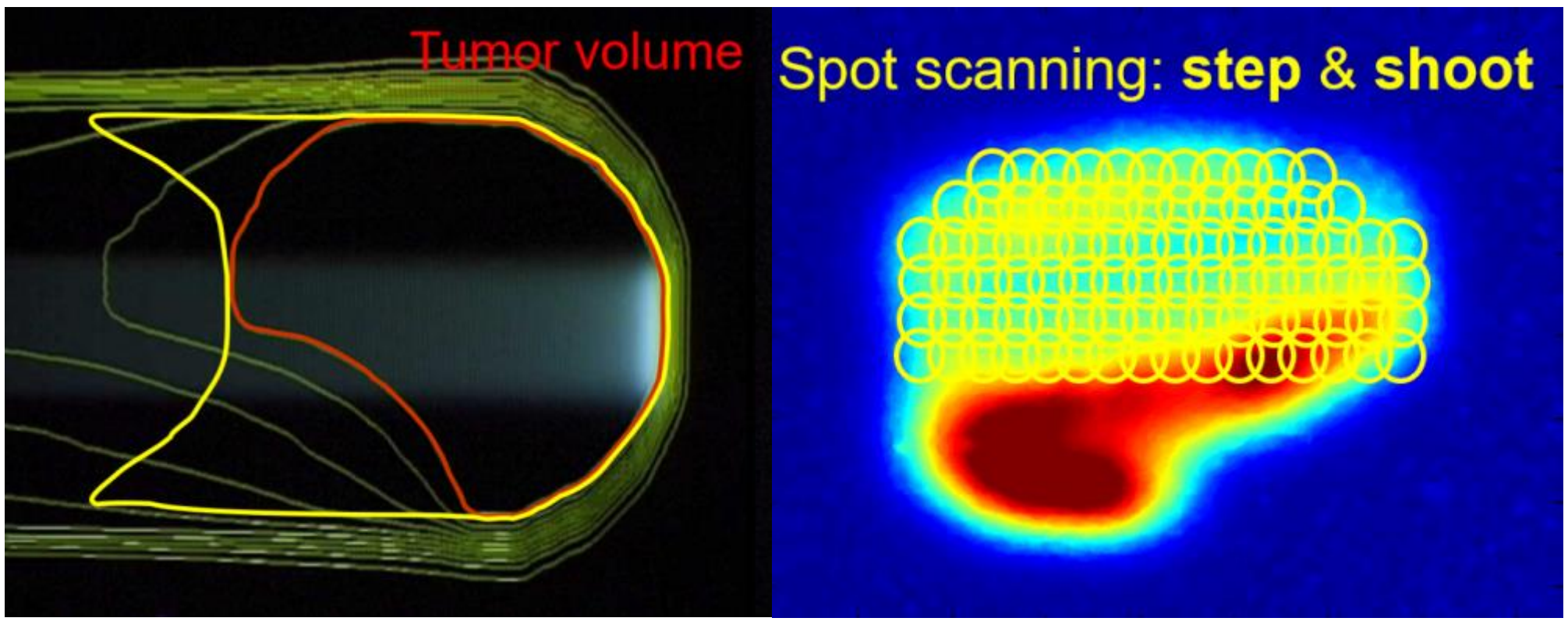

**Fig. 4:** Layer painting with a scanned pencil beam, shown in side view (left) and beam's-eye view (right).

### 4.4 Upstream and downstream scanning

Scanning magnets can be installed upstream or downstream of the final bend. In upstream scanning, the deflected beam passes through the final bending system. This requires a large magnet aperture, which increases cost and weight. The advantage is a nearly parallel beam at the patient, corresponding to an effectively infinite source-to-axis distance. The entrance field does not shrink as much relative to the field at depth.

In downstream scanning, the scanning magnets are placed after the final bend. The final bend then transports only the nominal trajectory, so the magnet aperture can be smaller. The scanned beam is divergent, giving a finite source-to-axis distance. This can increase the entrance dose and make the relationship between monitor measurements, air gap, and dose distribution more geometry dependent. Both concepts are clinically usable. The choice is an optical and engineering trade-off. The two scanning layouts are compared in Fig. 5.

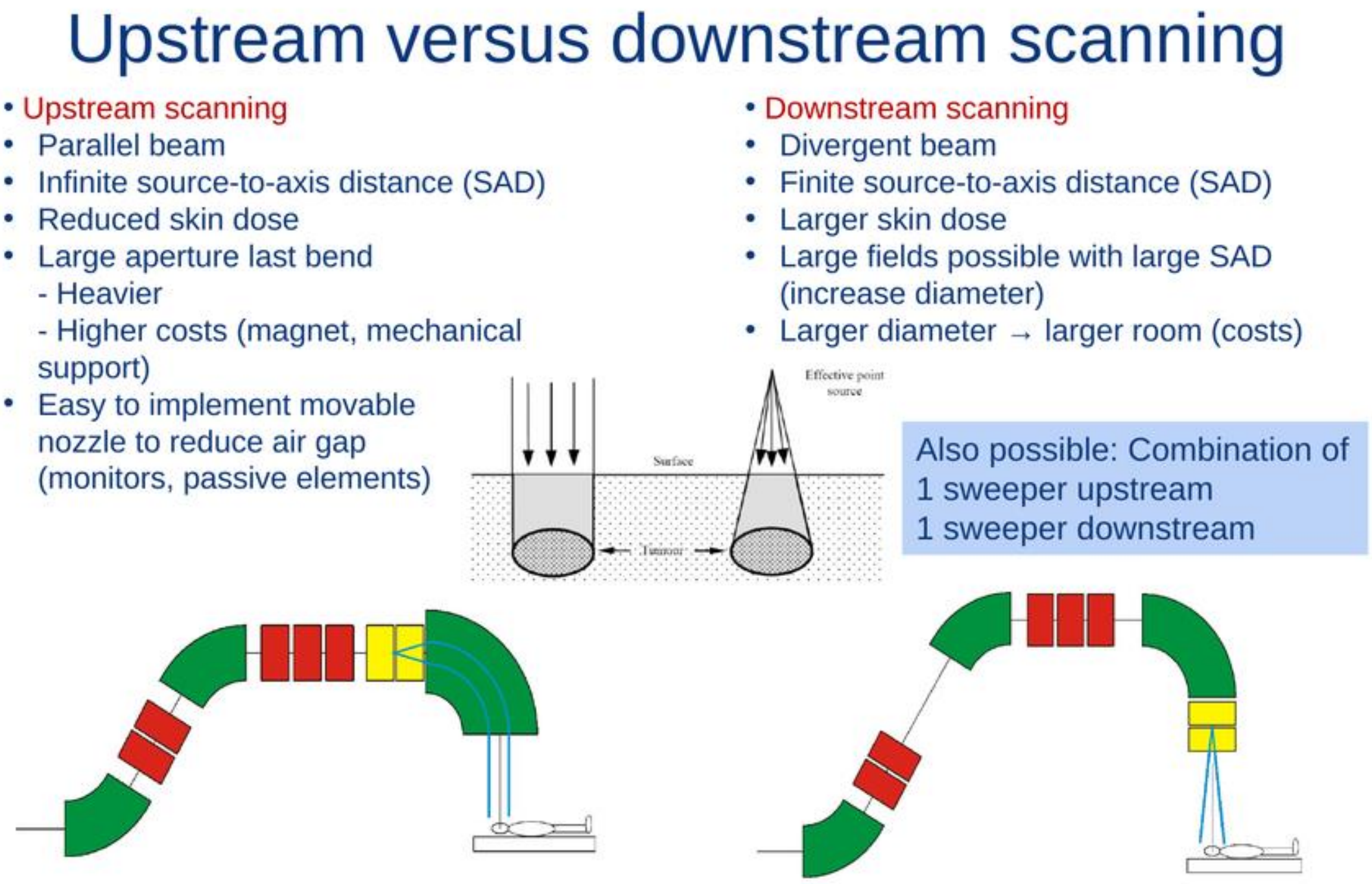

**Fig. 5:** Comparison of upstream and downstream scanning layouts in a proton gantry.

## 5 Reducing irradiation time

Modern proton therapy centres are hospital systems rather than research installations. They require compactness, high reliability, ease of maintenance, and short treatment interruptions. Treatment speed is also clinically relevant. Shorter irradiation times improve patient comfort, increase throughput, reduce sensitivity to organ motion, and facilitate repeated repainting of moving targets.

### 5.1 Motion mitigation and rescanning

Respiratory motion is a major concern for thoracic and abdominal targets. Cardiac motion and anatomical changes such as bladder or rectal filling can also displace selected targets by millimetres or centimetres. These displacements are comparable to or larger than the pencil-beam spot size. Motion mitigation strategies include gating, tumour tracking, and rescanning [9,10].

Gating delivers the beam only during a selected phase of the breathing cycle. It improves geometric accuracy but reduces duty cycle. Tracking attempts to follow the tumour in real time by changing the scanned beam position, but it requires reliable motion monitoring and low latency. Rescanning delivers the same volume multiple times. If the prescribed fluence is divided among N equal-weight paintings and their random dose-error patterns are statistically independent, the standard deviation of the random component decreases as

$$\sigma_{\mathrm{rand,N}} = \sigma_{\mathrm{rand,1}} / \sqrt{N}. \tag{6}$$

Here N is the total number of paintings. Correlated interplay patterns and systematic anatomical or range errors do not follow this scaling. Volumetric rescanning requires fast scanning in x and y and rapid changes in beam energy.

### 5.2 Continuous scanning and intensity modulation

Delivery time can be reduced by avoiding a stop at every spot. In continuous or TV-mode scanning, the beam moves through the target while its intensity is modulated. At PSI Gantry 2, for example, the beam current is modulated with a response time of about 100 μs during continuous line scanning [7]. The delivered dose is controlled by synchronized scanning magnet motion, beam-current modulation, and dose monitoring. Such modes can reduce dead time and support fast volumetric repainting.

### 5.3 Large-momentum-acceptance transport

For volumetric rescanning, energy switching is often the limiting step. A degrader or accelerator may be able to change energy quickly, but conventional beamline magnets must also be retuned to the new momentum. Large-momentum-acceptance beamlines address this bottleneck by transporting a wide momentum range without changing the magnetic field.

Fixed-field alternating-gradient optics use strong focusing cells to keep orbit excursions small over a large momentum interval. A published alternating-gradient canted-cosine-theta (AG-CCT) gantry design accepts a total momentum interval of about 25%, from -5% to +20% [12]. Alternatively, only the limiting bend can be designed for large acceptance. Combined-function superconducting magnets can provide bending, focusing, and higher-order corrections in one element. A momentum acceptance of around ±12% corresponds locally to a much larger range acceptance of approximately ±40% around a selected depth [11].

Moderate momentum acceptance is already clinically useful. The beamline does not need to accept the full 70 to 230 MeV range in one setting. If the required SOBP fits within the momentum window, the whole field can be delivered without changing the bend field. Larger targets can often be divided into two depth subfields. Large acceptance can therefore avoid many gantry-magnet retuning operations associated with energy-layer switching, although the individual beam energies must still be produced by the accelerator or degrader [11]. The number of required bend-field settings for the range and SOBP-width distribution used in that study is shown in Fig. 6.

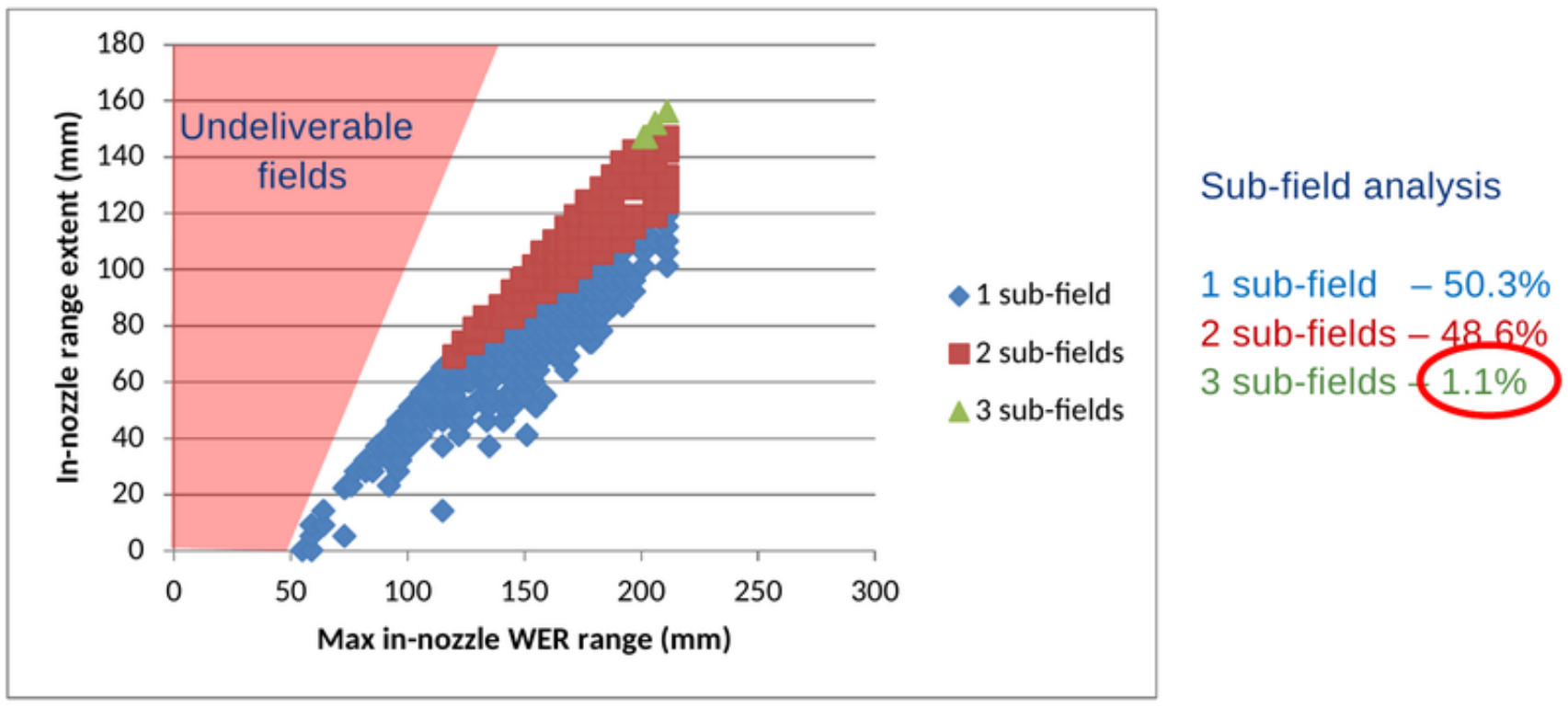

**Fig. 6:** Number of bend-field settings required with ±12% momentum acceptance, shown against the maximum in-nozzle water-equivalent range and the longitudinal range extent of each field.

## 6 Summary

Proton therapy relies on the Bragg peak, but the quality of treatment depends on the beam delivery system. Beamlines must transport the beam while controlling emittance, dispersion, imaging, and chromatic effects. In fixed-energy cyclotron- and synchrocyclotron-based systems, which remain the most common accelerator class in the current clinical inventory, the degrader and energy selection system set the treatment energy and define the accepted momentum spread. This process is robust and clinically effective, but it introduces emittance growth, beam loss, and secondary radiation.

Transverse dose delivery can be achieved by passive scattering or pencil-beam scanning. Passive scattering is conceptually simple but places substantial material in the beam path. Pencil-beam scanning keeps the beam narrow, reduces unnecessary material, and enables intensity-modulated dose delivery. Current developments focus on speed. Fast scanning, volumetric rescanning, and large-momentum-acceptance transport aim to reduce irradiation time and make proton therapy more robust for moving targets.

## Acknowledgements

The author thanks D. Meer and M. Schippers for teaching material, discussions, and permission to adapt selected teaching figures.

## References

[1] H. Paganetti (Ed.), Proton Therapy Physics, CRC Press, Boca Raton, 2012.

[2] NIST, PSTAR stopping power and range tables for protons, https://physics.nist.gov/PhysRefData/Star/Text/PSTAR.html.

[3] J.M. Schippers and A.J. Lomax, Emerging technologies in proton therapy, Acta Oncol. 50 (2011) 838-850, https://doi.org/10.3109/0284186X.2011.582513.

[4] M.J. van Goethem et al., Geant4 simulations of proton beam transport through a carbon or beryllium degrader and following a beam line, Phys. Med. Biol. 54 (2009) 5831-5846, https://doi.org/10.1088/0031-9155/54/19/011.

[5] Z. Liang et al., Design of the energy selection system for proton therapy, Proc. Cyclotrons 2016, MOD01.

[6] Particle Data Group, Passage of particles through matter, Prog. Theor. Exp. Phys. 2020 (2020) 083C01.

[7] S. Safai, C. Bula, D. Meer and E. Pedroni, Improving the precision and performance of proton pencil beam scanning, Transl. Cancer Res. 1 (2012) 196-206.

[8] E. Pedroni et al., The PSI Gantry 2: a second generation proton scanning gantry, Z. Med. Phys. 14 (2004) 25-34, https://doi.org/10.1078/0939-3889-00194.

[9] C. Bert and M. Durante, Motion in radiotherapy: particle therapy, Phys. Med. Biol. 56 (2011) R113-R144, https://doi.org/10.1088/0031-9155/56/16/R01.

[10] J.M. Pakela et al., Management of motion and anatomical variations in charged particle therapy: past, present, and into the future, Front. Oncol. 12 (2022) 806153, https://doi.org/10.3389/fonc.2022.806153.

[11] A. Gerbershagen et al., A novel beam optics concept in a particle therapy gantry utilizing the advantages of superconducting magnets, Z. Med. Phys. 26 (2016) 224-237, https://doi.org/10.1016/j.zemedi.2016.03.006.

[12] W. Wan et al., Alternating-gradient canted cosine theta superconducting magnets for future compact proton gantries, Phys. Rev. ST Accel. Beams 18 (2015) 103501, https://doi.org/10.1103/PhysRevSTAB.18.103501.